\documentclass[prd,nofootinbib,11pt]{revtex4}
\usepackage{amssymb,amsmath,latexsym,braket,fancyref,hyphenat}

\numberwithin{equation}{section}

\begin{document}

\begin{flushright}
CERN-PH-TH/2016-057\\ 
MAN/HEP/2016/03\\
March 2016
\end{flushright}

\title{{\Large Symmetries for SM Alignment in multi-Higgs Doublet Models}\\[3mm] }

\author {\large Apostolos Pilaftsis$^{\,a,b}\,$\footnote{E-mail address: {\tt apostolos.pilaftsis@manchester.ac.uk}}\\}

\affiliation{\vspace{0.2cm} 
${}^a$Consortium for Fundamental Physics, School
  of Physics and Astronomy, University of Manchester, Manchester, M13
  9PL, United Kingdom
${}$\vspace{2mm}\\
${}^b$Theory Division, CERN, CH-1211 Geneva 23, Switzerland}

\begin{abstract}
\vspace{2mm}
\centerline{\small {\bf ABSTRACT} }
\vspace{2mm}\noindent
We derive the complete set of continuous maximal symmetries for Standard Model
(SM) alignment  that may occur  in the tree-level scalar  potential of
multi-Higgs Doublet Models, with  $n >  2$ Higgs doublets.   Our results
generalize the symmetries of  SM alignment, without decoupling
of  large  mass scales  or  fine-tuning,  previously obtained  in  the
context of two-Higgs Doublet Models.

\end{abstract}

\maketitle

\section{Introduction}

As more  and more data  are being collected  at the CERN  Large Hadron
Collider~(LHC),   it   becomes   even  more   apparent   from   global
analyses~\cite{ATLAS:2014kua,Khachatryan:2014jya,Cheung:2014noa,
  Chowdhury:2015yja,Craig:2015jba} that the  observed Higgs boson $h$,
with a  mass $M_h  \approx 125$~GeV, interacts  with the  gauge bosons
with  coupling strengths  that are  very close  to those  predicted by
the~SM~\cite{Agashe:2014kda}.   The   constraints  deduced   from  the
strengths  of these  Higgs  couplings, primarily  to  $W^\pm$ and  $Z$
bosons,  put severe  limits on  the actual  form of  a possible  heavy
scalar sector  in the observable  sub-TeV range,  and so on  the model
structure of New Physics to be anticipated at the~LHC.

If there are  additional heavy scalars in the theory,  as predicted in
two-Higgs                        Doublet                      Models
(2HDMs)~\cite{Lee:1973iz,Pilaftsis:1999qt,review}  or   $n$HDMs,  with
$n                    >                    2$                    Higgs
doublets~\cite{Weinberg:1976hu,Grzadkowski:2010au,Felipe:2014zka,
  Keus:2014isa,Branco:2015bfb,Emmanuel-Costa:2016vej},  there are  two
main strategies that are followed in the literature to avoid too large
mixings of the heavy scalars to the SM Higgs boson~$h$.  The first one
is         known         as        the         {\em         decoupling
  limit}~\cite{Georgi:1978ri,Gunion:2002zf,Ginzburg:2004vp},  in which
the  heavy scalars  are made  very heavy,  such that  they effectively
decouple  from  the  low-energy  SM  sector  altogether.   The  second
strategy, which is of interest to us,  is a bit more optimistic, as it
leaves  open  the  possibility  to directly  probe  the  heavy  scalar
sector. It assumes that the new  scalars are not very heavy after all,
e.g.~they  have  masses in  the  sub-TeV  range, but  the  theoretical
parameters are arranged in such a way  that the new scalars do not mix
significantly  with~$h$.  As  a consequence,  the $hW^+W^-$  and $hZZ$
couplings retain  their SM values,  to a very good  approximation.  In
the 2HDM,  such parameter arrangement  leads to a sort  of `alignment'
between the  two-by-two CP-odd  scalar mass matrix~${\cal  M}^2_P$ and
the  two-by-two  CP-even  scalar  mass matrix~${\cal  M}^2_S$  in  the
CP-conserving limit of the theory.  In the so-called SM {\em alignment
  limit}~\cite{Ginzburg:1999fb,
  Chankowski:2000an,Ginzburg:2001ss,Delgado:2013zfa,Carena:2013ooa,
  Dev:2014yca,Bernon:2015qea,Bernon:2015wef},     the     two     mass
matrices~${\cal M}^2_P$  and~${\cal M}^2_S$ get diagonalized  by means
of the same mixing angle, e.g.~$\beta$, where $\tan\beta = v_2/v_1$ is
the ratio  of the vacuum  expectation values  (VEVs) of the  two Higgs
doublets              $\Phi_1$              and              $\Phi_2$,
i.e.~$\langle       \Phi_1\rangle       =      v_1/\sqrt{2}$       
and~$\langle \Phi_2\rangle = v_2/\sqrt{2}$.

An unpleasant aspect of achieving  SM alignment without decoupling of
mass scales is  the degree of fine-tuning that  is frequently required
among  the  theoretical  parameters.   However,  a  recent  study  has
shown~\cite{Dev:2014yca}  that  the  phenomeno\-logically  desirable  SM
alignment  in  the  2HDM  may  be  realized  upon  the  imposition  of
symmetries  on  its scalar  potential.  In  particular, three  maximal
symmetries, softly broken  possibly by bilinear mass  terms, have been
identified  for which  the  tree-level scalar  potential  of the  2HDM
exhibits exact SM alignment.   We therefore refer to this  mechanism as the
{\em  natural  alignment}  mechanism.  We  briefly  review  the  basic
features of natural alignment in the 2HDM in Section~\ref{sec:2HDM}.

In  this paper  we  extend  the results,  previously  obtained in  the
context of  2HDMs, and derive  the complete  set of symmetries  for SM
alignment that may take place in  multi-HDMs, with more than two Higgs
doublets.   In  $n$HDMs,  with  $n>2$  Higgs  doublets,  the  task  of
identifying all  SM alignment  symmetries becomes more  laborious.  An
$n$HDM may  have a number  $m$ of inert  Higgs doublets, which  do not
participate in the mechanism  of Electro-Weak Symmetry Breaking~(EWSB)
because    of   some    unbroken   symmetry,    such   as    a   $Z_2$
symmetry~\cite{Glashow:1976nt,    Deshpande:1977rw,   Silveira:1985rk,
  Barbieri:2006dq}, resulting in  the vanishing of VEVs  for all inert
Higgs doublets.  Nevertheless, the key  observation to be made here is
that  SM  alignment  is  naturally  achieved  in  multi-HDMs,  if  the
dimension-4 part  of the  scalar potential~$V$ containing  the quartic
couplings  of  the EWSB  scalars  is  invariant under  rotations  that
diagonalize  their corresponding  squared mass  matrix of  dimension-2
in~$V$.  To put it simply,  the quartic couplings involving {\em only}
EWSB   scalars   must   be   invariant   under   SM-alignment-symmetry
transformations    which   include    the    so-called   {\em    Higgs
  basis}~\cite{Georgi:1978ri}  in which  the bilinear  mass matrix  is
diagonal. Our approach to deriving  the symmetries for SM alignment in
multi-HDMs is presented in Section~\ref{sec:multiHDM}.

The  paper  has  the  following structure.   After  this  introductory
section, Section~\ref{sec:2HDM} briefly reviews  the basic features of
natural alignment  in the 2HDM.   After having gained  useful insight
from  the  2HDM  case,  we derive  in  Section~\ref{sec:multiHDM}  the
complete set  of symmetries  for SM  alignment that  may occur  in the
tree-level scalar potential of  multi-HDMs.  Finally, the key findings
of our study are summarized in Section~\ref{sec:Conclusions}.

\section{Natural Alignment in the 2HDM}\label{sec:2HDM}

In this section we briefly review the SM alignment limit in the 2HDM and 
the symmetries that naturally enforce this limit. More details may be
found in~\cite{Dev:2014yca}. 

Let  us start  our discussion  by writing  down the  tree-level scalar
potential $V$  of the 2HDM, expressed  in terms of the SU(2)$_L$ Higgs doublets
$\Phi_1$ and $\Phi_2$ with hypercharges $Y_{\Phi_{1,2}} = 1$,
\begin{eqnarray}
  \label{eq:V2HDM}
V \!& = &\! -\:\mu_1^2 (\Phi_1^{\dagger} \Phi_1) - \mu_2^2
(\Phi_2^{\dagger} \Phi_2) - m_{12}^2 (\Phi_1^{\dagger} \Phi_2) -
m_{12}^{*2}(\Phi_2^{\dagger} \Phi_1) + \lambda_1 (\Phi_1^{\dagger}\Phi_1)^2
 + \lambda_2 (\Phi_2^{\dagger} \Phi_2)^2\nonumber\\
\!&&\! +\: \lambda_3 (\Phi_1^{\dagger}
\Phi_1)(\Phi_2^{\dagger} \Phi_2) + \lambda_4 (\Phi_1^{\dagger}
\Phi_2)(\Phi_2^{\dagger} \Phi_1)  + \frac{\lambda_5}{2} (\Phi_1^{\dagger} \Phi_2)^2 +
\frac{\lambda_5^{*}}{2} (\Phi_2^{\dagger} \Phi_1)^2\\\
\!&&\! +\: \lambda_6 (\Phi_1^{\dagger} \Phi_1) (\Phi_1^{\dagger} \Phi_2) + \lambda_6^{*}
(\Phi_1^{\dagger} \Phi_1)(\Phi_2^{\dagger} \Phi_1) + 
\lambda_7 (\Phi_2^{\dagger} \Phi_2) (\Phi_1^{\dagger} \Phi_2) +
\lambda_7^{*} (\Phi_2^{\dagger} \Phi_2) (\Phi_2^{\dagger} \Phi_1)\; .\nonumber
\end{eqnarray}
Note that $V$  contains 4 real mass  parameters, $\mu_1^2$, $\mu_2^2$,
${\rm Re}\, m_{12}^2$  and ${\rm Im}\, m^2_{12}$, and  10 real quartic
couplings,  $\lambda_{1,2,3,4}$,  ${\rm   Re}\,  \lambda_{5,6,7}$  and
${\rm Im}\, \lambda_{5,6,7}$. To account for the mechanism of EWSB, 
the Higgs doublets $\Phi_{1,2}$  are expressed as
\begin{equation}
   \label{eq:Phi12}
\Phi_1\ =\ \left( \begin{array}{c}\phi^+_1 \\ \frac{1}{\sqrt{2}} ( v_1
                                              + \phi_1 + ia_1 )\end{array} \right)\; ,\qquad
\Phi_2\ =\ e^{i\xi}\, \left( \begin{array}{c}\phi^+_2 \\ \frac{1}{\sqrt{2}} ( v_2
                                              + \phi_2 + ia_2 )\end{array} \right)\; ,
\end{equation}
where $\xi$  is a CP-odd  phase, $\phi_{1,2}$ ($a_{1,2}$)  are CP-even
(CP-odd)  scalars  in  the  CP-conserving limit  of  the  theory,  and
$\phi^+_{1,2}$   are   positively   charged  scalars   in   the   weak
basis. Focusing on  the neutral scalar sector,  the corresponding mass
matrix in the  basis $( a_1\,, a_2\,, \phi_1\,, \phi_2)$  takes on the
symmetric form:
\begin{equation}
  \label{eq:MN}
{\cal M}^2_N \ =\  \left( \begin{array}{cc}
{\cal M}^2_P & {\cal M}^2_{PS} \\[2mm]
({\cal M}^{2}_{PS})^{\sf T} & {\cal M}^2_S \end{array}\right)\; ,
\end{equation}
where     ${\cal     M}^2_P     =     ({\cal     M}^2_P)^{\sf     T}$,
${\cal M}^2_S  = ({\cal  M}^2_S)^{\sf T}$ and  $ {\cal  M}^2_{PS}$ are
$2\times   2$   matrices   whose    explicit   form   may   be   found
in~\cite{Pilaftsis:1999qt}. Notice that in  the CP-conserving limit of
the theory,  all CP-violating  scalar-pseudoscalar mass  terms vanish,
i.e.~$ {\cal M}^2_{PS}  = 0$.  Then, in this  CP-conserving limit, the
$2\times 2$  mass matrices  ${\cal M}^2_P$ and  ${\cal M}^2_S$  can be
diagonalized separately by the SO(2) orthogonal transformations,
\begin{equation}
   \label{eq:OTs}
{\cal  \widehat{M}}^2_P \ = \ \left( \begin{array}{cc}
c_\beta & s_\beta \\
-s_\beta  & c_\beta\end{array}\right)\,
{\cal     M}^2_P\,
\left( \begin{array}{cc}
c_\beta & -s_\beta \\
s_\beta  & c_\beta\end{array} \right)\; ,\qquad
{\cal  \widehat{M}}^2_S \ = \ \left( \begin{array}{cc}
c_\alpha & s_\alpha \\
-s_\alpha  & c_\alpha\end{array}\right)\,
{\cal     M}^2_S\,
\left( \begin{array}{cc}
c_\alpha & -s_\alpha \\
s_\alpha  & c_\alpha\end{array} \right)\; ,
\end{equation}
with   $s_x  \equiv   \sin   x$   and  $c_x   \equiv   \cos  x$,   and
$x  = \alpha\,,  \beta$.  Upon  diagonalization, the  weak states  are
related to the mass eigenstates through:
\begin{equation}
   \label{eq:eigen}
\left( \begin{array}{c} a_1 \\ a_2 \end{array} \right)\ =\ 
\left( \begin{array}{cc}
c_\beta & -s_\beta \\
s_\beta  & c_\beta\end{array} \right)\, \left( \begin{array}{c} G^0 \\
                                                 A \end{array}
                                             \right)\;, \qquad
\left( \begin{array}{c} \phi_1 \\ \phi_2 \end{array} \right)\ =\ 
\left( \begin{array}{cc}
c_\alpha & -s_\alpha \\
s_\alpha  & c_\alpha\end{array} \right)\, \left( \begin{array}{c} h \\
                                                 H \end{array}
                                             \right)\;,
\end{equation}
where  $G^0$  is the  would-be  Goldstone  boson associated  with  the
longitudinal  degree  of polarization  of  the  $Z$  boson, $A$  is  a
physical CP-odd  scalar, $h$ is  identified with the  observed CP-even
Higgs boson at the  LHC, and $H$ is a new  heavy CP-even scalar having
mass  $M_H  >  M_h$~\footnote{It is straightforward to extend our 
discussion to scenarios with an inverted hierarchy: $M_H < M_h$.}.

According to  our conventions, the  SM alignment  limit of the  2HDM is
defined  as  the   limit  $\alpha  \to  \beta$.    As  was  explicitly
demonstrated in~\cite{Dev:2014yca},  SM alignment  is realized  in the
2HDM, {\em iff} the following condition is satisfied:
\begin{equation}
   \label{eq:align2HDM}
\lambda_7\, t_\beta^4\: -\: (2\lambda_2-\lambda_{345})\,t_\beta^3 \: +\:
3(\lambda_6-\lambda_7)\,t_\beta^2 \: +  \:
(2\lambda_1-\lambda_{345})\,t_\beta \: -\: \lambda_6 \ = \ 0 \; ,
\end{equation}
with  $\lambda_{345} \equiv  \lambda_3 +  \lambda_4 +  \lambda_5$, for
finite values  of $\tan\beta  \equiv t_\beta$, upon  implementation of
the      SM      Higgs     mass      constraint~\cite{Carena:2013ooa}:
$M_h \approx  125$~GeV.  Obviously,  if $\lambda _6  = 0$,  one simple
solution   to~(\ref{eq:align2HDM})   for   having  SM   alignment   is
$t_\beta =  0$.  By analogy, if  $\lambda_7 = 0$, SM  alignment may be
achieved       for      infinite       values      of       $t_\beta$,
namely~when~$t_\beta \to \infty$.

In the so-called SM alignment  limit $\alpha \to \beta$, the $h$-boson
coupling  to $W^\pm$  and  $Z$  bosons has  exactly  the SM  strength,
whereas the heavier  $H$ boson does not interact with  the $W^\pm$ and
$Z$  bosons   at  the  tree-level,  i.e.~it   becomes  partially  {\em
  gaugophobic}~\footnote{We should clarify  that the  $H$ boson  is not
  entirely gaugophobic. Although the trilinear couplings $HW^+W^-$ and
  $HZZ$  are  absent  in  the SM  alignment  limit,  quadrilinear
  interactions, such as $HHW^+W^-$  and $HHZZ$,  are  non-zero.}.  In
fact, such a  scenario is getting increasingly favourable in the
light of global  analyses of LHC data. Nevertheless,  the SM alignment
condition~(\ref{eq:align2HDM}) would  require an unpleasant  degree of
fine-tuning among the quartic couplings, unless there is some symmetry
that enforces it.  As a first  step to identify possible symmetries of
SM  alignment,  we  follow~\cite{Dev:2014yca}  and  require  that  the
condition~(\ref{eq:align2HDM})  is fulfilled  for {\em  any} value  of
$t_\beta$. Imposing this constraint, we find that
\begin{eqnarray}
   \label{eq:alcond}
\lambda_1 \ = \ \lambda_2 \ = \ \lambda_{345}/2\;, \qquad \lambda_6 \
  = \ \lambda_7 \ = \ 0\; .  
\end{eqnarray}
It is crucial  to observe here that in deriving  the above constraint,
no reference  was made  on the  structure of  the bilinear  mass terms
$\mu^2_{1,2}$     and    $m^2_{12}$     of    the     potential    $V$
in~(\ref{eq:V2HDM}). In principle, their  form is not restricted from
SM-alignment considerations, as  long as they lead  to a CP-conserving
theory. Their role is simply to fix the parameter $\tan\beta$ to a particular value,
thus relating the weak basis $(\Phi_1\,,\Phi_2)$ to the Higgs (mass-eigenstate) basis.

The  next step  taken in~\cite{Dev:2014yca}  was to  identify possible
maximal symmetries  of the 2HDM  potential~$V$ that would  enforce the
constraint~(\ref{eq:alcond}).  The complete  classification of all the
13  SU(2)$_L$-invariant  maximal  symmetries  of  the  2HDM  has  been
presented  in~\cite{Battye:2011jj,Pilaftsis:2011ed},  after  extending
the          U(1)$_Y$-restricted           bilinear          formalism
of~\cite{Maniatis:2007vn,Ivanov:2007de,Nishi:2007dv,  Ferreira:2009wh,
  Ferreira:2010yh}. In this way, the following three symmetries for SM
alignment  have been  identified~\cite{Dev:2014yca}~\footnote{With the
  aid   of    the   bilinear    formalism,   the    three   symmetries
  in~(\ref{eq:2HDM3syms})          are         classified          as:
  (i)~${\rm        SO(5)}        \simeq       {\rm        Sp}(4)/Z_2$;
  (ii)~${\rm        SO(3)}       \simeq        {\rm       SU}(2)_{\rm HF}/Z_2$;
  (iii)~${\rm SO}(2)\times Z_2 \simeq {\rm SO}(2)_{\rm HF}\times {\cal CP}$,
  e.g.~according to Table~2 of~\cite{Pilaftsis:2011ed}}:
\begin{eqnarray}
   \label{eq:2HDM3syms}
\mbox{(i)}   &&  
           {\rm Sp(4)}: \qquad\qquad\qquad  \lambda_1 \ = \ \lambda_2 \ = \
                \lambda_3/2\;, \quad\hspace{2.5mm}
\lambda_4\ =\ \lambda_5 \ = \ \lambda_6 \ = \ \lambda_7\ = \ 0\;,\nonumber\\
\mbox{(ii)}   && 
           {\rm SU(2)_{\rm HF}}: \qquad\qquad\ \;  \lambda_1 \ = \
                 \lambda_2 \ = \ \lambda_{34}/2\;, \quad\ \lambda_5 \ = \
                 \lambda_6 \ = \ \lambda_7\ = \ 0\;, \\ 
\mbox{(iii)}  && {\rm SO(2)_{\rm HF}}\times {\cal CP}: \qquad
                 \lambda_1 \ = \ \lambda_2 \ = \ \lambda_{345}/2\;,
                 \quad\hspace{-0.3mm} \lambda_6 \ = \ \lambda_7\ = \ 0 \; .\nonumber
\end{eqnarray}  
We note  that the  unitary symplectic group  Sp(4)~\footnote{We remind
  the reader about  the basic set relations of  ${\rm Sp}(2n)$ groups:
  ${\rm  SU}(n)  \subset  {\rm  Sp}(2n) \subset  {\rm  SU}(2n)$,  with
  $n\ge 2$.}  is  acting on the reduced  four-dimensional ${\bf \Phi}$
basis,                            defined                           as
${\bf  \Phi}  \equiv  (   \Phi_1\,,  \Phi_2\,,  i\sigma^2  \Phi^*_1\,,
i\sigma^2\Phi^*_2                      )^{\sf                     T}$,
where  $\sigma^2$  is  the  second   matrix  of  the  Pauli  matrices:
$\sigma^a  = (\sigma^1\,,  \sigma^2\,, \sigma^3)$.   Hence, the  group
Sp(4)   defines   a   larger   set   of   {\em   custodial}   symmetry
transformations~\cite{Pilaftsis:2011ed}.  As for  the symmetries SU(2)
in~(ii) and  SO(2) in~(iii), these  are acting on  the two-dimensional
Higgs  Family~(HF) space:  $(\Phi_1\,, \Phi_2)$.   Therefore, we  also
denote   these    symmetries   as    ${\rm   SU(2)_{\rm    HF}}$   and
${\rm  SO(2)_{\rm HF}}$,  respectively.  Finally,  the discrete  group
${\cal   CP}$  in~(iii)   refers   to  the   canonical  CP   symmetry:
$\big(\Phi_1(t,   {\bf   x})\,,   \Phi_2(t,   {\bf   x})   \big)   \to
\big(\Phi^*_1(t,   -{\bf    x})\,,   \Phi^*_2(t,    -{\bf   x})\big)$,
which  is tacitly  assumed to  apply to  the classical  action of  the
theory.

Having  gained valuable  insight from  the above  exercise, one  might
wonder  how   the  three  symmetries   of  the  SM   alignment  stated
in~(\ref{eq:2HDM3syms}),  Sp(4),  SU(2)   and  SO(2),  would  manifest
themselves   in  an   explicit   construction  of   the  2HDM   scalar
potential~$V$. To  this end,  we observe  that there  are correspondingly
three symmetry structures that are relevant to SM alignment:
\begin{eqnarray}
  \label{eq:SDaT}
S &=& \Phi^\dagger_1 \Phi_1\: +\: \Phi^\dagger_2 \Phi_2\ =\
      \frac{1}{2}\, {\bf \Phi}^\dagger\,{\bf \Phi}\; ,\nonumber\\
D^a &=& \Phi^\dagger_1 \sigma^a \Phi_1\: +\: \Phi^\dagger_2\sigma^a\Phi_2\; ,\\[2mm]
T &=& \Phi_1\Phi_1^{\sf T} \: +\: \Phi_2\Phi_2^{\sf T} \; .\nonumber
\end{eqnarray}
Under         an        SU(2)$_L$         gauge        transformation:
$\Phi_{1,2}\:  \to\:   \Phi'_{1,2}  \,   =\,  U\,   \Phi_{1,2}$,  with
$U  \in {\rm  SU}(2)_L$, the  symmetry structures  $S$, $D^a$  and $T$
transform as follows:
\begin{equation}
   \label{eq:SDaTSU2L}
S\ \to\ S'\ =\ S\;, \qquad D^a\ \to\ D'^a\ =\ O^{ab} D^b\;, \qquad 
T\ \to\ T'\ =\ U\, T\, U^{\sf T}\; , 
\end{equation}
where $O \in {\rm SO(3)}$. Evidently, $S$ is
a gauge-invariant  scalar, $D^a$ transforms  as a 3D  Euclidean vector
and  $T$  transforms  as  a  bi-doublet.   Under  Higgs-doublet  field
transformations,  the  quantity  $S$  is  an  Sp(4)-invariant  in  the
${\bf \Phi}$-space,  whilst $D^a$ and  $T$ are invariant under  SU(2) and
SO(2) rotations in the HF space $(\Phi_1\,, \Phi_2)$, respectively.

In terms  of $S$, $D^a$  and $T$, the  most general scalar 2HDM
potential $V$ realizing  natural alignment may alternatively be
written down as follows:
\begin{equation}
  \label{eq:V2hdm}
V\ =\  V_{\rm sym} \: +\: \Delta V\; ,
\end{equation}
where
\begin{eqnarray}
   \label{eq:potSDT}
V_{\rm sym} & = & -\,\mu^2\,S\: +\: \lambda_S\, S^2\: +\: \lambda_D\,
        D^aD^a\: +\: \lambda_T\, {\rm Tr}\,(T\,T^*)\nonumber\\
&=&  -\,\mu^2\, (\Phi^\dagger_1   \Phi_1  +  \Phi^\dagger_2  \Phi_2) \:
    +\: (\lambda_S +\lambda_D + \lambda_T)\, \Big[
    (\Phi^\dagger_1\Phi_1)^2  + 
        (\Phi^\dagger_2\Phi_2)^2\Big]\\
&& +\: 2\,(\lambda_S -\lambda_D )\,
        (\Phi^\dagger_1\Phi_1)\, (\Phi^\dagger_2\Phi_2)
\: +\:  4\lambda_D\, (\Phi^\dagger_1\Phi_2)(\Phi^\dagger_2\Phi_1)\:
   +\:  \lambda_T\Big[ (\Phi^\dagger_1\Phi_2)^2 +
   (\Phi^\dagger_2\Phi_1)^2 \Big]\; ,\nonumber 
\end{eqnarray}
with $\mu^2 > 0$~\footnote{In addition to the condition for EWSB,
  convexity of $V$ at large values of $\Phi_{1,2}$ in {\em any} field
  direction can be simply enforced by demanding that
  $\lambda_{S,D,T} \ge 0$.}, and
\begin{equation}
   \label{eq:DV2hdm}
\Delta V\ = \  \sum\limits_{i,j =1,2} m^2_{ij}\, \Phi^\dagger_i\Phi_j
\end{equation}
are soft-symmetry  breaking terms.  In  arriving at the  last equality
in~(\ref{eq:potSDT}),  we  have employed  the  property  of the  Pauli
matrices:
~$(\Phi^\dagger_1 \sigma^a  \Phi_1)\,(\Phi^\dagger_2 \sigma^a \Phi_2)\
=\     2\,(\Phi^\dagger_1\Phi_2)(\Phi^\dagger_2\Phi_1)      \,     -\,
(\Phi^\dagger_1\Phi_1)\,                      (\Phi^\dagger_2\Phi_2)$.
Obviously, the  so-called Maximally  Symmetric 2HDM studied  in detail
in~\cite{Dev:2014yca},     which    corresponds     to    Scenario~(i)
of~(\ref{eq:2HDM3syms}),  is obtained  when  $\lambda_D =  \lambda_T =  0$,
whilst    the   SU(2)$_{\rm    HF}$-symmetric   2HDM    (Scenario~(ii)
of~(\ref{eq:2HDM3syms}))  is  recovered  when  $\lambda_T  =  0$.  Finally,
Scenario~(iii) of~(\ref{eq:2HDM3syms})  is realized,  if all  three quartic
couplings $\lambda_{S,D,T}$ in~(\ref{eq:potSDT}) are non-zero.

We   now    notice   that   the   SM    alignment   symmetries   given
in~(\ref{eq:2HDM3syms})  can be  used to  diagonalize the  dimension-2
part of the  2HDM potential~$V$ in the HF  space $(\Phi_1\,, \Phi_2)$.
This means that the Hermitian bilinear mass matrix, which we define in
the HF space as $M^2_{ij} \equiv -\mu^2\delta_{ij} + m^2_{ij}$, can be
brought  into the  diagonal form  $\widehat{M}^2_{ij}$ by  means of  a
SU(2)  transformation, {\em  without}  altering  the dimension-4  part
of~$V$,   i.e.~$V_{\rm   sym}$    given   in~(\ref{eq:potSDT}).    For
Scenario~(iii), the bilinear mass matrix $M^2_{ij}$ must be real to be
diagonalizable     by     means      of     an     SO(2)     rotation,
i.e.~$M^2_{ij}   =   M^{2\,*}_{ij}$.    In    this   HF   basis,   say
$(\Phi'_1\,,\Phi'_2)$, in which $M^2_{ij}$ is diagonal, only one Higgs
doublet  acquires non-zero  VEV, e.g.~$\Phi'_1$,  which is  identified
with   the  VEV   $v$  of   the   SM  Higgs   doublet  field   $\Phi$,
i.e.~$\langle   \Phi'_1  \rangle   \equiv  \langle   \Phi  \rangle   =
v/\sqrt{2}$.
This specific HF  basis, i.e.~$(\Phi'_1\,, \Phi'_2)$, is  also called the
{\em Higgs basis}~\cite{Georgi:1978ri}.  Hence, the key observation is
that SM-alignment-symmetry transformations not  only leave the quartic
couplings invariant, but also include transformations that lead to the
{\em Higgs  basis}. Notice  that in  the Higgs  basis, an  {\em exact}
canonical $Z_2$ symmetry  for the 2HDM potential  becomes manifest, in
which $\Phi'_1 \to +\Phi'_1$  and $\Phi'_2  \to -\,\Phi'_2$,  which remains
unbroken,  even after  EWSB, i.e.~$\Phi'_2$  becomes an  inert doublet
(see  also   our  discussion   below  for   the  inert   2HDM).   This
symmetry-based  approach  will  prove  very  useful  in  deriving  the
complete set  of symmetries for  alignment in $n$HDMs, with  $n>2$, in
the next section.

Besides   the   softly   broken   SM   alignment   symmetries   stated
in~(\ref{eq:2HDM3syms}), under which the  EWSB Higgs doublets $\Phi_1$
and~$\Phi_2$  have  non-trivial   transformation  properties,  another
possible way for  getting natural alignment in 2HDM will  be to impose
an    unbroken     discrete    $Z_2$    symmetry~\cite{Glashow:1976nt,
  Deshpande:1977rw,Silveira:1985rk,Barbieri:2006dq},  under which  one
of the  Higgs doublets  is $Z_2$-odd,  e.g.~$\Phi_2 \to  -\Phi_2$.  In
this class  of scenarios,  in which  $\lambda_6 =  \lambda_7 =  0$ and
$m^2_{12} = 0$, whilst the quartic couplings $\lambda_{1,2,3,4,5}$ are
unconstrained, the  Higgs basis  is fixed to  be along  the $Z_2$-even
scalar field, e.g.~$\Phi_1$, representing the SM Higgs doublet~$\Phi$.
Such naturally aligned scenarios, in which the $Z_2$-odd Higgs doublet
is an  {\em inert} field,  correspond to  the limits: $t_\beta  \to 0$
(for   $\Phi_1   \equiv   \Phi$)   or  $t_\beta   \to   \infty$   (for
$\Phi_2 \equiv \Phi$).  As we shall see in the next section, the inert
scalar sector in multi-HDMs can  have much richer structure possessing
its own set of symmetries.

\section{Natural Alignment in multi-HDMs}\label{sec:multiHDM}

It  is  straightforward to  generalize  the  results of  the  previous
section and  derive the  complete set of  symmetries for  alignment in
$n$HDMs, with  $n>2$.  We  may assume  that the  scalar sector  of the
theory consists of $m$ inert Higgs doublets $\widehat{\Phi}_{\hat{a}}$
(with  $\hat{a}  =  \hat{1},\,  \hat{2},\dots,  \hat{m}$),  for  which
$\langle  \widehat{\Phi}_{\hat{a}}  \rangle  =  0$,  and  $N_H \equiv n-m$  Higgs
doublets $\Phi_a$  (with $a  = 1,\,  2,\dots, N_H$)  which generally
take part in electroweak symmetry  breaking (EWSB) with non-zero VEVs,
i.e.~$\langle \Phi_a \rangle \neq 0$. The vanishing of the VEVs of the
inert Higgs  doublets is enforced  by imposing a suitable  discrete or
continuous  symmetry ${\cal  D}$, which  remains unbroken  after EWSB.
Under  the action  of  ${\cal  D}$, the  {\em  non}-inert $N_H$  Higgs
doublets $\Phi_a$  transform trivially, i.e.~$\Phi_a \to  \Phi_a$.  In
general,  the discrete  group ${\cal  D}$ could  be $Z_2$,  or even  a
higher $Z_N$ symmetry, but {\em  not} the canonical ${\cal CP}$ and/or
a permutation  symmetry, such as~$S_3$,  since such symmetries  do not
necessarily  imply the  vanishing  of  the VEVs  for  the inert  Higgs
doublets $\widehat{\Phi}_{\hat{a}}$.

The full  scalar potential $V$  of a  naturally aligned $n$HDM may be
written down as a sum of three terms:
\begin{equation}
  \label{eq:Vmulti}
V \ =\  V_{\rm sym}\: +\: V_{\rm inert}\: +\: \Delta V\; ,
\end{equation}
where  $V_{\rm sym}$  describes the  symmetry-constrained part  of the
scalar sector,  which is responsible for  the EWSB of the  theory. Its
form turns out to be  the same as the  one   found   in   the
2HDM~[cf.~\eqref{eq:potSDT}], which we will elucidate  in  more  detail
below.   In    addition,   the   potential   term    $V_{\rm   inert}$
in~(\ref{eq:Vmulti}) represents the inert scalar sector which does not
participate into EWSB, i.e.
\begin{eqnarray}
  \label{eq:Vinert}
V_{\rm inert} & = & \widehat{m}^2_{\hat{a}\hat{b}}\,
                    \widehat{\Phi}^\dagger_{\hat{a}}
                    \widehat{\Phi}_{\hat{b}}
\: +\: \lambda_{\hat{a}\hat{b}\hat{c}\hat{d}}\, (\widehat{\Phi}^\dagger_{\hat{a}}
                    \widehat{\Phi}_{\hat{b}}) (\widehat{\Phi}^\dagger_{\hat{c}}
                    \widehat{\Phi}_{\hat{d}}) \: +\: 
\lambda_{\hat{a}\hat{b}c d}\, (\widehat{\Phi}^\dagger_{\hat{a}}
                    \widehat{\Phi}_{\hat{b}}) (\Phi^\dagger_c\Phi_d)\:
                    +\:  
\lambda_{a\hat{b}\hat{c}d}\, (\Phi^\dagger_a \widehat{\Phi}_{\hat{b}}) 
   (\widehat{\Phi}^\dagger_{\hat{c}}\Phi_d) 
                    \nonumber\\[2mm]
&&+\:
\Big[\,\lambda_{a\hat{b}c\hat{d}}\,
                    (\Phi^\dagger_a\widehat{\Phi}_{\hat{b}})
                    (\Phi^\dagger_c \widehat{\Phi}_{\hat{d}} )\: +\:
{\rm H.c.}\Big]\; ,
\end{eqnarray}
where summation of  the indices over their allowed range  of values is
understood,        and       $\lambda_{\hat{a}\hat{b}\hat{c}\hat{d}}$,
$\lambda_{\hat{a}\hat{b}c   d}$,  $\lambda_{a\hat{b} \hat{c} d}$  and
$\lambda_{\hat{a}b\hat{c}d}$  are   all  different  sets   of  quartic
couplings. The $m\times  m$ matrix $\widehat{m}^2_{\hat{a}\hat{b}}$ is
taken to  be positive  definite, so as to avoid  spontaneous~EWSB.   From the
explicit construction of the inert scalar sector in~(\ref{eq:Vinert}),
we readily see that it remains invariant under the $Z_2$ symmetry,
\begin{equation}
   \label{eq:Z2inertV}
Z^{\rm I}_2:\quad \Phi_a\ \to \ \Phi_a\;, \qquad \widehat{\Phi}_{\hat{b}}\ \to\ -\,
\widehat{\Phi}_{\hat{b}}\; .
\end{equation}
Consequently,  $Z^{\rm I}_2$ should  always  be contained  in  the ${\cal  D}$
symmetry      group     of      the      inert     scalar      sector,
i.e.~$Z^{\rm I}_2 \subset {\cal D}$.  Finally,  the  third term  $\Delta  V$  in~(\ref{eq:Vmulti})
contains the soft-symmetry breaking mass parameters of the EWSB sector
and is given by
\begin{equation}
   \label{eq:DeltaV}
\Delta V\ = \  m^2_{ab}\, \Phi^\dagger_a\Phi_b\; ,
\end{equation}
where $m^2_{ab}$ is in general a Hermitian $N_H\times N_H$ matrix.
Note   that   dimension-2   mixed    bilinear   operators,   such   as
$\widehat{\Phi}^\dagger_{\hat{a}}  \Phi_b$,  are  not allowed  in  the
theory.

On the  basis of the  assumption that the soft-symmetry  breaking mass
matrix~$m^2_{ab}$  in~\eqref{eq:DeltaV}  has  no  particular  discrete
symmetry structure~\footnote{Our assumption  that~$m^2_{ab}$ must have
  no particular discrete symmetry structure may also be motivated from
  the strong constraints  on the non-observation of  domain walls that
  are  produced from  the  spontaneous breaking  of possible  ungauged
  discrete  symmetries  in the  theory  at  the SM  electroweak  phase
  transition.},  there are  then  three  possible continuous  symmetry
groups that can  bring~$m^2_{ab}$ into the diagonal  form, as required
to  happen  in  the  Higgs  basis.   These  three  continuous  maximal
symmetries for  SM alignment,  which should be  respected by  the EWSB
potential   term~$V_{\rm    sym}$   in~\eqref{eq:Vmulti},    are   the
generalization of those found in the 2HDM, i.e.
\begin{eqnarray}
  \label{eq:symMulti2HDM}
({\rm i}) ~~{\rm Sp}(2N_H) \times {\cal D}\,,\qquad 
({\rm ii}) ~~{\rm SU}(N_H) \times {\cal D}\,,\qquad 
({\rm iii}) ~~{\rm SO}(N_H) \times {\cal CP} \times {\cal D}\; ,
\end{eqnarray}
where  the  discrete  ${\cal  CP}$  symmetry is  acting  on  the  {\em
  non}-inert  scalar   sector  consisting  of  $N_H = n-m$   Higgs  doublets
$\Phi_a$, with $N_H > 1$. We note  that, for $N_H = 1$, the EWSB
part  of the  potential  $V_{\rm  sym}$ becomes  identical  to the  SM
potential, whereas the  symmetry group ${\cal D}$ of  the inert sector
may have  its own  rich structure,  as we  will see below.  As
mentioned in  Section~\ref{sec:2HDM}, we emphasize again  that for the
maximal  symmetry~(iii),   the  soft-symmetry  breaking   mass  matrix
$m^2_{ab}$ must be real or close to  real, in order to avoid too large
tree-level CP-violating scalar-pseudoscalar transitions, whose effects
are   severely    constrained   by    limits   on    electric   dipole
moments~\cite{Cheung:2014oaa,Dekens:2014jka}.

We  note  that  as  in  the  2HDM, the  effect  of  $\Delta  V$  given
in~(\ref{eq:DeltaV})  will  be  to  fix the  rotation  angles  of  the
continuous symmetries in~(\ref{eq:symMulti2HDM})  to particular values
that  relate  the  weak-basis  fields $\Phi_a$  to  the  corresponding
fields~$\Phi'_a$ in  the Higgs (mass-eigenstate) basis.   In the Higgs
basis,  one of  the rotated  scalar doublets,  e.g.~$\Phi'_1$, becomes
aligned to  the SM Higgs  doublet $\Phi$, i.e.~$\Phi'_1  \equiv \Phi$,
such                                                              that
$\langle \Phi'_1 \rangle  = \langle \Phi \rangle =  v/\sqrt{2}$ is the
SM VEV.   Like in the 2HDM case, in the  Higgs basis, an  {\em exact}
canonical $Z_2$ symmetry for the the EWSB part of the $n$HDM potential
becomes manifest,
\begin{equation}
     \label{eq:Z2EW}
Z^{\rm EW}_2:\quad \Phi'_1 \to \Phi'_1\;,\qquad \Phi'_{a'} \to -\,\Phi'_{a'}\;,
\end{equation}
where $a' = 2,3,\dots, N_H$.  As a consequence, after EWSB, the
full      $n$HDM      potential      becomes      invariant      under
$Z^{\rm  EW}_2\!\times  Z^{\rm I}_2$,  where  $Z^{\rm  I}_2$ is  given
in~(\ref{eq:Z2inertV}).   In  fact,  this  residual  product  symmetry
$Z^{\rm EW}_2\!\times Z^{\rm  I}_2$ is instrumental, as it  is the one
that enforces SM  alignment in the $n$HDM, even  beyond the tree-level
approximation.

Let us now turn our attention  to the symmetry-constrained part of the
scalar sector  $V_{\rm sym}$ that  occurs in the $n$HDM  potential $V$
in~\eqref{eq:Vmulti}. We will demonstrate  that its analytical form is
the same  as the one  found in the 2HDM.  To elucidate this  point, we
first introduce the symmetry-covariant structures:
\begin{equation}
   \label{eq:SDTmulti}
S_{\cal A} \ =\ \sum\limits_{k,l= 1}^{N_H}\Phi^\dagger_k\, [{\cal
  A}_{kl}\otimes {\bf 1}_2]\, \Phi_l\; ,\qquad
D_{\cal B}^a\ =\ \sum\limits_{k,l = 1}^{N_H} \Phi^\dagger_k\, [{\cal
  B}_{kl}\otimes \sigma^a]\, \Phi_l\; ,\qquad
T_{\cal C} \ =\ \sum\limits_{k,l = 1}^{N_H} \Phi_k\,{\cal C}_{kl}\,\Phi_l^{\sf T} \; ,
\end{equation}
where ${\cal A} = {\cal A}^\dagger$, ${\cal B} = {\cal B}^\dagger$ and
${\cal C} = \pm\, {\cal C}^{\sf  T}$ are all $N_H\times N_H$ matrices.
Observe   that   the   three   symmetry   structures   $S_{\cal   A}$,
$D_{\cal  B}^a$ and  $T_{\cal C}$  all share  the same  transformation
properties under the  SU(2)$_L$ gauge group as  the corresponding ones
given in~(\ref{eq:SDaTSU2L}) for the 2HDM.  Specifically, $S_{\cal A}$
is  a  gauge-invariant singlet,  $D_{\cal  B}^a$  transforms as  a  3D
Euclidean vector,  and $T_{\cal  C}$ transforms  as a  bi-doublet.  In
principle,  one could  have  also considered  more involved  covariant
objects                  of                 the                  form:
$D_{\cal B}^{ab\dots}  = \sum_{k,l = 1}^{n-m}  \Phi^\dagger_k\, [{\cal
  B}_{kl}\otimes      (\sigma^a\sigma^b\cdots)]       \,      \Phi_l$.
Then,  gauge-invariant objects,  such  as $(D_{\cal  B}^{ab\dots})^2$,
which would  potentially contribute  to $V_{\rm  sym}$, can  always be
reduced  to  $(S_{\cal  A})^2$  and $(D_{\cal  B}^a)^2$,  through  the
successive          use           of          the          identities:
$\sigma^a \sigma^b = {\bf  1}_2\delta^{ab} + i\epsilon^{abc} \sigma^c$
and $\epsilon^{abc}\epsilon^{abd} = 2 \delta^{cd}$. Hence, such higher
rank  tensors, under  the SU(2)$_L$  gauge group,  do not  introduce new
candidate structures for $V_{\rm sym}$, other than the ones stated in~\eqref{eq:SDTmulti}.

Let us  now consider symmetry transformations  in the HF space  of the
EWSB sector. Under such  transformations, the structures~$S_{\cal A}$,
$D_{\cal  B}^a$ and  $T_{\cal C}$  stated in~(\ref{eq:SDTmulti})  must
transform         covariantly          with         respect         to
${\rm  SU}(N_H) \subset  {\rm  Sp}(2N_H)$ and  ${\rm SO}(N_H)$  global
groups.   This means  that the  $N_H\times N_H$  matrices ${\cal  A}$,
${\cal B}$ and ${\cal C}$ must  be products of the generators~$T^a$ of
either  ${\rm  SU}(N_H)$  or  ${\rm SO}(N_H)$  groups  (including  the
identity ${\bf  1}_{N_H}$), in  the fundamental  representation. These
products of~$T^a$  generate tensors  in the group  space and  may also
have  particular   symmetry-orderings  as  determined  by   the  Young
tableaux.   For instance,  non-trivial  rank-2 tensor  objects can  be
produced,                                                           if
${\cal A}\,,\ {\cal B}\,,\ {\cal C} = -i\, [T^a\,,\ T^b] = f^{abc}\,T^c$,
or                                                                  if
${\cal A}\,,\ {\cal  B}\,,\ {\cal C} = \{T^a\,,  T^b\} \propto d^{abc}
T^c$, for ${\rm SU}(N_H> 2)$ and ${\rm SO}(N_H > 4)$. However, taking
into account the known identities for the groups generators,
\begin{eqnarray}
\mbox{SU($N_H$)}: && (T^a)_{ij} (T^a)_{kl} \ =\  \frac{1}{2}\, \Big(
                 \delta_{il}\,\delta_{kj}\: -\: \frac{1}{N_H}\,\delta_{ij}\,\delta_{kl}\Big)\;,\\
\mbox{SO($N_H$)}: && (T^a)_{ij} (T^a)_{kl} \ =\  2\, \Big(
                 \delta_{il}\,\delta_{kj}\: -\: \delta_{ij}\,\delta_{kl}\Big)\;,
\end{eqnarray}
and                   the                  fact                   that
$f^{abc}f^{abd},\, d^{abc}d^{abd} \propto \delta^{cd}$, it is then not
difficult  to show  that all  symmetry-invariant objects  of potential
interest,   such  as   $(S_{\cal   A})^2$,   $(D_{\cal  B}^a)^2$   and
${\rm Tr} (T_{\cal C}T^*_{\cal C})$,  reduce to linear combinations of
the    objects     $(S_{\bf    1})^2$,    $(D_{\bf     1}^a)^2$    and
${\rm      Tr}      (T_{\bf      1}     T^*_{\bf      1})$,      with
${\cal   A}   =    {\cal   B}   =   {\cal   C}    =   {\bf   1}_{N_H}$
in~\eqref{eq:SDTmulti}.    Therefore,   the   HF-singlet   structures,
$S_{\bf  1}$, $D_{\bf  1}^a$ and  $T_{\bf 1}$,  will be  sufficient to
describe   the  symmetry-constrained   part  of   the  scalar   sector
$V_{\rm sym}$, i.e.
\begin{equation}
  \label{eq:Vsym}
V_{\rm sym}\ =\ -\, \mu^2 S_{\bf 1}\: +\: \lambda_S\, S^2_{\bf 1}\: +\: 
\lambda_D\, D_{\bf 1}^a D_{\bf 1}^a \: +\: \lambda_T\, {\rm Tr}\, (
T_{\bf 1}\,T^*_{\bf 1})\;,  
\end{equation} 
with $\mu^2  > 0$. Consequently,  the general group-structure  form of
$V_{\rm sym}$ in $n$HDMs (with $n > 2$) will be the same as the  one found in the
2HDM~[cf.~\eqref{eq:potSDT}].

For      illustration,     let      us  now   apply      our     findings
in~(\ref{eq:symMulti2HDM}),  in order  to  obtain  all the  admissible
forms for naturally aligned 2HDM and 3HDM potentials.  In the 2HDM, if
both   the   Higgs   doublets  $\Phi_{1,2}$   participate   in   EWSB,
(\ref{eq:symMulti2HDM}) leads obviously to  the three maximal symmetry
groups  presented  in Section~\ref{sec:2HDM}:  (i)~Sp(4);  (ii)~SU(2);
(iii)~SO(2)$\times {\cal CP}$ [cf.~(\ref{eq:2HDM3syms})].
Now,    if     one    of     the   scalar    doublets of the 2HDM   is    inert,
e.g.~$\widehat{\Phi}_{\hat{2}}$  ($m=1$), then  there are  {\em three}
distinct possibilities:
\begin{itemize}

\item[(i)] the  $Z_2 = Z^{\rm I}_2$ symmetry~\cite{Glashow:1976nt, Deshpande:1977rw,
    Silveira:1985rk,  Barbieri:2006dq} as  discussed  in the  previous
  section [cf.~(\ref{eq:Z2inertV})];

\item[(ii)] the {\em unbroken} Peccei--Quinn-type U(1)
  symmetry~\cite{Peccei:1977hh,Peccei:1977ur}, enforcing that 
  $\lambda_{1\hat{2}1\hat{2}} \equiv \lambda_5 = 0$, under which
  $\widehat{\Phi}_{\hat{2}}$ is charged;

\item[(iii)]             the            custodial             symmetry
  ${\rm Sp(2)}  \simeq {\rm  SU(2)}_{C,{\rm I}}$  acting on  the {\em inert}
  field                                                          space
  $(\widehat{\Phi}_{\hat{2}}\,, i\sigma^2\widehat{\Phi}^*_{\hat{2}})$,
  which implies that $\lambda_{1\hat{2}1\hat{2}} \equiv \lambda_5 = 0$
  {\em and}  $\lambda_{1\hat{2}\hat{2}1} \equiv  \lambda_4 =  0$. This
  symmetry               is                classified               as
  ${\rm SO(4)} \simeq {\rm SU(2)}_{C,{\rm EW}} \times {\rm SU(2)}_{\rm
    C,                            {\rm                            I}}$
  in  Table~1~(No.~11)  of~\cite{Pilaftsis:2011ed},  where  the  first
  custodial group ${\rm SU(2)}_{C,{\rm EW}}$ acts on the EWSB sector.

\end{itemize}

We now turn our attention to  the 3HDM. If all three Higgs doublets
participate in the  mechanism of EWSB which corresponds  to $m=0$, the
symmetries of alignment resulting from~(\ref{eq:symMulti2HDM}) are:
\begin{equation}
  \label{eq:3HDM0}
({\rm i}) ~~{\rm Sp}(6)\,,\qquad 
({\rm ii}) ~~{\rm SU}(3)\,,\qquad 
({\rm iii}) ~~{\rm SO}(3) \times {\cal CP}\; ,
\end{equation} 
where  Sp(6)  is acting  on  the  six-dimensional ${\bf  \Phi}$-multiplet:
${\bf  \Phi}   \equiv  (   \Phi_1\,,  \Phi_2\,,   \Phi_3\,,  i\sigma^2
\Phi^*_1\,,   i\sigma^2\Phi^*_2\,,   i\sigma^2\Phi^*_3   )^{\sf   T}$,
and SU(3) and  SO(3) on the HF space:  $(\Phi_1\,, \Phi_2\,,\Phi_3 )$.
This   means  that   the  3HDM   potential  has   the  symmetry   form
of~(\ref{eq:Vsym}),    softly    broken     by    terms    as    given
in~(\ref{eq:DeltaV}).

Considering   now    the   case   of   one    inert   Higgs   doublet,
e.g.~$\widehat{\Phi}_{\hat{3}}$ ($m=1$),  the symmetries of  the inert
sector     will      be:     (i)~the     $Z_2$      symmetry,     with
$\widehat{\Phi}_{\hat{3}}  \to   -\widehat{\Phi}_{\hat{3}}$;  (ii)~the
global U(1)  symmetry, under which  $\widehat{\Phi}_{\hat{3}}$ carries
non-zero  charge, leading  to $\lambda_{a\hat{3}b\hat{3}}  = 0$  (with
$a,b     =    1,2$);     (iii)~the     custodial    symmetry     group
${\rm  Sp(2)}  \simeq  {\rm  SU(2)}_C$   acting  on  the  field  space
$(\widehat{\Phi}_{\hat{3}}\,,   i\sigma^2\widehat{\Phi}^*_{\hat{3}})$,
which  implies   that  $\lambda_{a\hat{3}\hat{3}b}  =  0$   {\em  and}
$\lambda_{a\hat{3}b\hat{3}} = 0$.

The last possible class of naturally aligned 3HDMs is the one that has
two    inert    Higgs   doublets,    $\widehat{\Phi}_{\hat{2}}$    and
$\widehat{\Phi}_{\hat{3}}$ ($m=2$).   In addition  to the  inert $Z_2$
symmetry     in~(\ref{eq:Z2inertV}):      $\Phi_1     \to     \Phi_1$,
$\widehat{\Phi}_{\hat{2}}    \to     -\widehat{\Phi}_{\hat{2}}$    and
$\widehat{\Phi}_{\hat{3}} \to -\widehat{\Phi}_{\hat{3}}$, one inherits
at     least     all     13      maximal     symmetries     of     the
2HDM~\cite{Battye:2011jj,Pilaftsis:2011ed}
when~$\lambda_{1\hat{a}\hat{b}1}  =  \lambda_{1\hat{a}1\hat{b}}  =  0$
(with  $\hat{a}, \hat{b}  =  \hat{2}, \hat{3}$)  in~(\ref{eq:Vinert}),
where   Sp(4)   is   the   largest  symmetry   group.    However,   if
$\lambda_{1\hat{a}\hat{b}1}$   and  $\lambda_{1\hat{a}1\hat{b}}$   are
non-zero, further  symmetries, which  include~$Z^{\rm I}_2$, may exist  in the
3HDM    potential     that    forbid~$\widehat{\Phi}_{\hat{2}}$    and
$\widehat{\Phi}_{\hat{3}}$ from developing a VEV.  A recent example is
the   generalized   CP   symmetry   of   order-4   observed   recently
in~\cite{Ivanov:2015mwl}.  It  should be stressed here  again that all
the symmetries of the 3HDM  inert sector must remain unbroken under
EWSB.

For $n$HDMs with $n > 3$,  the complexity of the classification of the
inert scalar sector increases, but results from lower $n$-cases become
crucial  in  our  approach   to  constructing  all  naturally  aligned
multi-HDMs. Instead, the alignment  symmetries for the non-inert sector
are   fully  specified   by   the  three   maximal  symmetries   given
in~(\ref{eq:symMulti2HDM}). If  other {\em inert} scalars  are present
in the theory, e.g.~singlets $S_i$ or triplets $\Delta_l$, our results
will still hold  true for the EWSB part~(\ref{eq:Vsym})  of the scalar
potential.   Thus, the  approach  presented here  is  not confined  to
multi-HDMs only, but  it can easily be  generalized to more
abstract scalar sectors.

\section{Conclusions}\label{sec:Conclusions}

We  have  derived  the  complete  set of  maximal  symmetries  for  SM
alignment that could take place  in the tree-level scalar potential of
the SM,  with $n \ge  2$ Higgs  doublets.  Our results  generalize the
symmetries  of SM  alignment, previously  obtained in  the context  of
two-Higgs Doublet Models  (the $n=2$ case~\cite{Dev:2014yca}), without
decoupling  of  large  mass  scales  or  recourse  to  specific  model
parameter arrangements.   For the  scalar sector participating  in the
EWSB mechanism, the  general symmetry conditions for  {\em natural} SM
alignment are given by  Equation~(\ref{eq:symMulti2HDM}), which is one
of the central results of this paper.

Another highlight  of our  study is  that the  inert scalar  sector of
multi-HDMs may have  its own rich symmetry  structure.  In particular,
within our symmetry-based  approach, we have found that  the 2HDM with
one inert Higgs doublet may exhibit three distinct maximal symmetries:
(i)~the  discrete $Z_2  = Z^{\rm  I}_2$ symmetry~\cite{Glashow:1976nt,
  Deshpande:1977rw,Silveira:1985rk,Barbieri:2006dq};   (ii)~the   U(1)
symmetry and  (iii)~the custodial symmetry~${\rm  SU(2)}_{C,{\rm I}}$,
which all  remain unbroken after  EWSB. In addition to  the frequently
considered  symmetry~(i),  it  would  be interesting  to  explore  the
phenomenological   and   cosmological   implications  of   the   inert
symmetries~(ii) and~(iii).

We should  stress again that the  mechanism of SM alignement  does not
get invalidated, even  to {\em all orders},  if soft-symmetry breaking
masses [cf.~(\ref{eq:DeltaV})] are  added to the theory.   As noted in
Section~\ref{sec:multiHDM} [cf.~(\ref{eq:Z2EW})],  this is  because of
the  presence  of an  {\em  exact}  $Z^{\rm EW}_2\!\times  Z^{\rm  I}_2$
symmetry in the $n$HDM potential,  which becomes manifest in the Higgs
basis~\cite{Georgi:1978ri}.   In  particular,  beyond  the  tree-level
approximation,  the   alignement  symmetries  are  preserved   by  the
SU(2)$_L$ gauge  interactions, whereas the hypercharge  U(1)$_Y$ group
only breaks: ${\rm  Sp}(2n) \to {\rm SU}(n)$, without  spoiling the SM
alignment.   Instead,  the  Yukawa   sector  of  the  theory  violates
explicitly the  alignment symmetries, and  so it could  sizeably break
the  SM  alignment beyond  the  tree  level.  However,  such  explicit
violations  are  highly  model-dependent,  and  for  a  given  flavour
structure   of   the  Yukawa   sector,   they   lead  to   predictable
deviations~\cite{Dev:2014yca} from  the SM  values of  the Higgs-boson
couplings to $W^\pm$  and $Z$ bosons that may  constrain the parameter
space of the theory. In this context,  it is useful to remark that the
effect  of  these  constraints  can  be  drastically  reduced  in  the
so-called    Yukawa-aligned    models~\cite{Pich:2009sp},    if    the
quark-Yukawa basis happens to be coincidentally aligned with the Higgs
basis of the $n$HDM.  In~conclusion,  given the existing strict limits
on non-SM deviations in the Higgs couplings to~$W^\pm$ and~$Z$ bosons,
Equation~(\ref{eq:symMulti2HDM}) provides  an important  constraint on
future  model-building of  multi-HDMs predicting  additional low-scale
scalars with masses being in the explorable sub-TeV range of the~LHC.

\bigskip

\section*{Acknowledgements} 
\vspace{-3mm}
\noindent
This work is supported in part by the Lancaster--Manchester--Sheffield
Consortium for Fundamental Physics, under STFC research grant:
ST/L000520/1.

\end{document}